# Detection of charge states of an InAs nanowire triple quantum dot with an integrated nanowire charge sensor

Weijie Li,[1,2] Jingwei Mu,[1,2] Shaoyun Huang,[1] Dong Pan,[3,4] Jianhua Zhao,[3,4] and H. Q. Xu[1,4,a)]

[1] Beijing Key Laboratory of Quantum Devices, Key Laboratory for the Physics and Chemistry of Nanodevices and Department of Electronics, Peking University, Beijing 100871, China

[2] Academy for Advanced Interdisciplinary Studies, Peking University, Beijing 100871, China

[3] State Key Laboratory of Superlattices and Microstructures, Institute of Semiconductors, Chinese Academy of Sciences, P.O. Box912, Beijing 100083, China

[4] Beijing Academy of Quantum Information Sciences, Beijing 100193, China

[a)] Electronic address: hqxu@pku.edu.cn

(Dated: December 10, 2020)

ABSTRACT

A linear triple quantum dot (TQD) integrated with a quantum dot (QD) charge sensor is realized. The TQD and the charge sensor are built from two adjacent InAs nanowires by fine finger gate technique. The charge state configurations of the nanowire TQD are studied by measurements of the direct transport signals of the TQD and by detection of the charge state transitions in the TQD via the nanowire QD sensor. Excellent agreements in the charge stability diagrams of the TQD obtained by the direct transport measurements and by the charge-state transition detection measurements are achieved. It is shown that the charge stability diagrams are featured by three groups of charge state transition lines of different slopes, corresponding to the changes in the electron occupation numbers of the three individual QDs in the TQD. It is also shown that the integrated nanowire QD sensor is highly sensitive and can detect the charge state transitions in the cases where the direct transport signals of the TQD are too weak to be measurable. Tuning to a regime, where all the three QDs in the TQD are close to be on resonance with the Fermi level of the source and drain reservoirs and co-existence of triple and quadruple points becomes possible, has also been demonstrated with the help of the charge sensor in the region where the direct transport signals of the TQD are hardly visible.

Semiconductor quantum dots (QDs) have been extensively investigated in a variety of material systems, such as GaAs/AlGaAs heterostructures[1-5], InAs and InSb nanowires[6-8], carbon nanotubes[9-13], and Si/SiGe heterostructures[14,15], for potential application in quantum information processing and computation[16]. Among all these semiconductor materials, QDs fabricated in InAs nanowires are of great interest due to the presence of small effective mass, large Landé g-factor and strong spin-orbit interaction in the materials[17-24]. Double QDs and triple QDs (TQDs) have been realized in InAs nanowires[6,25-28] and have been used to demonstrate the feasibility in fast, all-electrical manipulations of electron spins[6]. Further studies of charge transfer events and charge rearrangements in multiple quantum dots (MQDs) in the few-electron regime inevitably require integration of charge readout elements[9,29-36]. However, limited by the one-dimensional geometry of nanowires, it is not straightforward to integrate a sensitive, conveniently tunable charge sensor with a MQD defined in an InAs nanowire[37-39]. Nevertheless, in a pioneer experiment, Hu *et al*. have developed an approach for charge sensing of a double QD defined in a Ge/Si core/shell nanowire by capacitively coupling of the double QD to a single QD on an adjacent nanowire via a floating metal gate[35]. Going to demonstrate integration of a sensitive charge sensor into an InAs nanowire MQD beyond a double QD is then a highly desired step towards the developments of InAs-nanowire based quantum logic gate circuits and quantum simulation devices.

In this letter, we report on integration of a QD charge sensor with a TQD defined all in InAs nanowires via fine finger gate technique. The coupling between the QD sensor and the TQD is achieved capacitively through a thin metal wire. The integrated device is studied by measurements of the charge stability diagrams of the TQD via the charge sensor and via direct electron transport through the TQD. We show that excellent agreements in the charge stability diagrams obtained by the two types of measurements are achieved. We also show that the charge sensor is highly sensitive and can detect the charge state transitions in the TQD in the regime where the direct transport signal is too weak to be detectable due to opaque tunnel couplings[40,41]. Finally, we demonstrate that it is achievable to tune the TQD into an energetically degenerated region, where all the three QDs in the TQD are close to be on resonance with the Fermi level of the source and drain reservoirs and the formation of quadruple points (QPs) becomes possible, with the assistance of the QD charge sensor.

The InAs nanowires employed in this work were grown via molecular-beam epitaxy

(MBE)[42]. For device fabrication, our MBE-grown InAs nanowires were transferred from the growth substrate onto a silicon substrate covered with a 300-nm thick layer of $SiO_2$. Pairs of adjacent InAs nanowires with a diameter of ~30 nm and a pure wurtzite crystalline phase were selected for subsequent fabrication processes. Within each pair, an InAs nanowire with a sufficient length was used to define a MQD and the other nanowire was used for fabrication of a QD charge sensor. The Ti/Au (5 nm/90 nm) source and drain contacts were fabricated onto both the MQD and charge sensor nanowires by standard electron-beam lithography (EBL), metal deposition via electron-beam evaporation (EBE) and lift off. Here we note that the contact regions of the nanowires were chemically etched in a $(NH_4)_2S_x$ solution to remove the surface oxide layer before the metal deposition. Then, after another step of EBL, a layer of $HfO_2$ with a thickness of 10 nm, covering the active nanowire areas of the devices, was fabricated by atomic layer deposition (ALD) and lift off. Finally, top finger gates together with coupling metal wires were fabricated by a third step of EBL, followed by EBE of 5-nm-thick Ti and 25-nm-thick Au, and lift off. Figure 1(a) shows a scanning electron microscope (SEM) image of a fabricated device, in which eleven top finger gates on the MQD nanowire labeled by G1 to G11, three top finger gates on the charge sensor nanowire labeled by g1 to g3, and the coupling metal wire labeled by CG are displayed. These top finger gates and the coupling wire have a width of ~30 nm, and the arrays on the MQD and charge sensor nanowires have a pitch of 60 nm and of 70 nm, respectively. Figure 1(b) shows a schematic cross-sectional view of the MQD nanowire in the device and a TQD defined in the MQD nanowire.

All electrical measurements presented here were performed in a $He^3/He^4$ dilution refrigerator at a base temperature of ~20 mK. Both the MQD and the charge sensor in each device were measured in a two-terminal DC setup with a bias voltage applied to the source and the drain contact being grounded. The Si substrate and the $SiO_2$ layer serve as a global back gate and a gate dielectric to the nanowires, respectively.

In this work, we present our measurements for a device similar to the one shown in Figure 1(a). In the measurements, we set the global back gate voltage at $V_{BG}$=8.8 V to keep the nanowires in a n-type conduction state. Before the measurements, the transfer characteristics of the finger gates were characterized and it was found that all these finger gates are capable of cutting off the nanowire current completely with threshold voltages applied in a range of −1 V to −0.3 V. In the measurements presented below, finger gates G1 to G4 and G11 were not used and were kept grounded.

Figure 2(a) shows the measured differential conductance $dI_{sd}/dV_{sd}$ of the single QD in the charge sensor as a function of source-drain voltage $V_{sd}$ and plunger gate voltage $V_{g2}$ (charge stability diagram). Here, the dot is defined by setting the gate voltages applied to gates g1 and g3 at $V_{g1}=-0.95$ V and $V_{g3}=-0.35$ V, and $I_{sd}$ is the source-drain current of the charge sensor. A Coulomb diamond pattern typical for a single QD is clearly observable in the measured charge stability diagram, indicating that single electron transport events occur through the QD in the charge sensor. From the measured Coulomb diamonds, the charging energy of the sensor QD is estimated to be $E_C \sim 4.5$ meV and the total capacitance of the sensor QD is extracted to be $C_\Sigma \sim 35$ aF. Figure 2(b) shows a typical charge stability diagram of a single QD defined in the MQD nanowire by using gates G7 and G9 as two barrier gates and G8 as a plunger gate. Here, $I_{SD}$ is the source-drain current of the MQD nanowire and $V_{SD}$ is the applied source-drain voltage. Similar charge stability diagrams were also obtained for single QDs defined in the MQD nanowire using any other three neighboring finger gates. Figure 2(c) shows a plot of the source-drain current $I_{sd}$ of the charge sensor as a function of $V_{g2}$ at $V_{sd}=0.1$ mV. Here, clear Coulomb current oscillations are observed. Operating the sensor QD at a steep slope of a Coulomb current peak enables sensitive detection of charge state transitions in a QD defined in the adjacent MQD nanowire. Figure 2(d) shows the simultaneous measurements of current $I_{sd}$ through the charge sensor at $V_{sd}=0.1$ mV and current $I_{SD}$ through the single QD defined by gates G7, G8 and G9 in the MQD nanowire as in Figure 2(b) at $V_{SD}=0.2$ mV. Here, the plunger-gate voltage $V_{g2}$ of the charge sensor was set at the declining slope of a Coulomb current peak, as indicated by a red star in Figure 2(c), and the measurements were performed as $V_{G8}$ is swept from 0.02 to 0.07 V. It is seen that distinct jumps in $I_{sd}$ appear when $I_{SD}$ is at Coulomb current peaks. Since sweeping $V_{G8}$ through a Coulomb oscillation peak of $I_{SD}$ leads to a change in electron occupation number of the single QD in the MQD nanowire, a jump in $I_{sd}$ corresponds to detection of a charge state transition in the single QD defined in the MQD nanowire.

Figure 3(a) shows the charge stability diagram of a TQD defined in the MQD nanowire obtained by the measurements of the direct electron transport current $I_{SD}$ through the TQD as a function of gate voltages $V_{G7}$ and $V_{G9}$ at $V_{SD} = 70$ μV. Here, the TQD is defined by gates G5, G7, G8 and G10 in the MQD nanowire as shown in Figure 1(b), where the three defined QDs are represented by blue ovals and labeled as QD1 QD2, and QD3. Note that QD2 is defined only by gates G7 and G8 as shown in Figure 1(b). Note also that voltages $V_{G7}$ and $V_{G9}$ applied to barrier gate G7 and plunger gate G9 were swept in the measurements for Figure 3(a), which allows us to efficiently map out the charge states of the three QDs in a two-dimensional diagram. Finite current lines of three different slopes can be observed in Figure 3(a). These finite current lines can be assigned to resonant transport through individual QD levels in the TQD. It should be pointed out that appearance of these resonant current lines also involves higher-order tunneling processes through other QDs which are not on resonance with the Fermi level of the source and drain electrodes. The couplings of the QD on resonance with the source and/or drain electrodes are achieved via these higher-order processes and are generally very small, making the resonant current lines very weak or hardly observable. In details, the current lines with a slope marked by inclined dashed lines arise from resonant transport through energy levels of QD2 in the TQD [see the schematic in the upper panel of Figure 3(b)]. This is due to the fact that both gate G7 and gate G9 couple to QD2 significantly. The current lines as marked by nearly vertical dashed lines arise from resonant transport through energy levels of QD1 in the TQD [see the schematic in the lower panel of Figure 3(b)]. This is because QD1 couples strongly only to gate G7, as gate G9 is located far from QD1. Similarly, the current lines as marked by nearly horizontal dashed lines arise from resonant transport through energy levels of QD3 in the TQD, since gate G7 is far from QD3 and couples only weakly to QD3. Figure 3(a) also shows that the inclined current lines are much brighter than the nearly vertical and nearly horizontal current lines. This is most likely due to the fact that the couplings of QD2 to the source and drain electrodes are more symmetric, since both couplings are achieved via higher-order tunneling processes though a single off-resonance QD. However, the couplings of QD1 or QD3 to the two reservoirs are largely asymmetric, which could lead to a strong suppression in tunneling current through the two QDs[9,30].

Figure 3(c) shows the charge stability diagram of the TQD detected by the QD charge sensor at setting $V_{g2}$=0.1 V (i.e., at the rising slope of a Coulomb current peak) and $V_{sd}$=0.2 mV, where the transconductance $dI_{sd}/dV_{G9}$ of the QD charge sensor was measured in the same ranges of $V_{G7}$ and $V_{G9}$ as in Figure 3(a). Dark transition lines of three different slopes (i.e., nearly vertical, inclined, and nearly horizontal lines) represent negative changes in $dI_{sd}/dV_{G9}$ and are thus the detection signals for the charge state transitions in the three individual QDs in the TQD. Compared to Figure 3(a), it is seen that the measured charge stability diagram by the QD charge sensor is in good agreement with that obtained by the direct transport measurements. However, it is important to note that the nearly vertical and the nearly horizontal charge state transition lines are clearly observed in Figure 3(c), while the corresponding current lines have been hardly visible in the direct transport measurements shown in Figure 3(a). Also it is interesting to see in Figure 3(c) that at anti-crossing regions where a horizontal dark line intersects with an inclined or with a vertical dark line, short bright detection lines, instead of dark lines, are observed. This is because at these anti-crossing regions, an increase in $V_{G9}$ will causes a charge redistribution among the three QDs with the center of electron charge moving away from the CG wire, leading to a decrease in electron charge detected by the sensor QD. The lengths of these bright lines reflect the combined strengths of capacitive and tunneling couplings between the two involved QDs. Thus, the capacitive and tunneling couplings between QD1 and QD3 are much weaker than that between QD2 and QD3, as can be discerned in Figure 3(c), which is consistent with the fact that the distance between QD1 and QD3 is larger than that between QD2 and QD3. Figure 3(d) shows the charge stability diagram of the TQD measured through the QD charge sensor by setting $V_{g2}$=0.114 V [i.e., at the declining slope of a Coulomb current peak as indicated by the red star in Figure 2 (c)]. Clearly, the same charge stability diagram as in Figure 3(c) is observed in Figure 3(d) although all the detected charge state transition lines are inverted in brightness (or darkness).

In a TQD, a quadruple point (QP) where all the three involved QDs are with energy levels on close resonance with the Fermi level of the source and drain reservoirs can be constructed for studies of various physical effects and for quantum information processing and simulations[2,3,29-33]. However, in a TQD with the QDs connected in series, searching for such a QP by the direct transport measurements is often difficult due to weak and invisible current in the measurements. Below, we show how such a QP can be found in our InAs nanowire TQD with use of the integrated QD charge sensor.

Figures 4(a)-4(c) show the charge stability diagrams of the same TQD as in Figure 3 measured as the transconductance $dI_{sd}/dV_{G9}$ of the QD charge sensor as a function of $V_{G7}$ and $V_{G9}$, at three different $V_{G6}$. Here, the charge sensor is again operated by setting $V_{g2}$=0.114 V and $V_{sd}$=0.2 mV [the same as in Figure 3(d)]. The TQD is initially set at a region with two triple points, where the energy levels of QD2 and QD3 are resonantly

aligned with the Fermi level of the source and drain reservoir, and $V_{G6}$ is tuned to bring the energy level of QD1 into resonance with the energy levels of QD2 and QD3. Figure 4(a) shows the charge stability diagram of the TQD detected by the charge sensor at $V_{G6}$ =−0.076 V. As compared to the stability diagram of the TQD at $V_{G6}$ =−0.072 V in Figure 3(d), the vertical charge state transition line of QD1 is seen to move closer to the region marked by a dashed circle in Figure 4(a), where two triple points are present. At this value of $V_{G6}$, the energy level of QD1 is still detuned from the energy levels of QD2 and QD3, as illustrated schematically in Figure 4(d). Figure 4(b) shows the charge stability diagram of the TQD at $V_{G6}$ =−0.081 V. Here, the three charge state transition lines are seen to locate closely and show a complex interaction pattern in the region marked by a dashed circle in Figure 4(b), where all the three QDs are close to be on resonance with the Fermi level of the source and drain reservoirs as illustrated schematically in Figure 4(e), and co-existence of a QP and several triple points can be discerned[30]. Figure 4(c) shows the detected charge stability diagram of the TQD at $V_{G6}$ =−0.084 V, where the region containing two triple points is marked by a dashed circle and the close resonance condition of all the three QDs is lifted as illustrated in Figure 4(f). Similar tuning procedures can be employed to tune the TQD into regions with different charge states in the three individual QDs, but again with co-existence of a QP and several triple points in the TQD (see Supplementary Material for such an example in a slightly stronger tunnel coupling case).

In summary, we have demonstrated detection of the charge state transitions in an InAs nanowire TQD using an integrated single-QD charge sensor defined in an adjacent nanowire. It is shown that the single-QD charge sensor is highly sensitive to detect charges states transitions in the TQD, allowing the measurements of the charge stability diagrams of the TQD in the regime where the direct transport signal through the TQD is invisible. The measured charge stability diagrams of the TQD are featured by three groups of charge state transition lines of different slopes with each group being associated with the charge state transitions in a QD in the TQD. We have also demonstrated a tuning of the TQD to a regime where all the three QDs are close to be on resonance with the Fermi level of the source and drain reservoirs and thus co-existence of a QP and several triple points becomes possible, with the help of the integrated QD charge sensor. We anticipate that the design and operation principles demonstrated in this work will facilitate developments of semiconductor-nanowire based quantum logic gates and quantum simulation processors.

SUPPLEMENTARY MATERIAL

See supplementary material for an example of tuning of the InAs nanowire TQD to a region where all the three QDs in the TQD are close to be on resonance with the source and drain electrodes in a slightly stronger coupling condition.

ACHNOWLEDGMENTS

This work was supported by the Ministry of Science and Technology of China through the National Key Research and Development Program of China (Grant Nos. 2017YFA0303304, 2016YFA0300601, 2017YFA0204901, and 2016YFA0300802), the National Natural Science Foundation of China (Grant Nos. 11874071, 11974030, 91221202, 91421303, and 61974138), the Beijing Academy of Quantum Information Sciences (Grant No. Y18G22), and the Beijing Natural Science Foundation (Grant No. 1202010). D.P. also acknowledges the support from the Youth Innovation Promotion Association of the Chinese Academy of Sciences (Grant No. 2017156).

DATA AVAILABILITY

The data that support the findings of this study are available within the article and supplementary material, and from the corresponding author upon request.

## CAPTIONS

**Figure 1.** (a) SEM image of a QD charge sensor integrated MQD device built from two adjacent InAs nanowires by fine top finger gate technique. Finger gates G1-G11 are fabricated on the MQD nanowire and finger gates g1-g3 are fabricated on the charge sensor nanowire. The charge sensor is coupled to the MQD via a thin metal wire labeled as CG in the image. (b) Cross-sectional schematic view of the MQD nanowire in the device, in which a TQD is defined and manipulated using gates G5-G10 (golden). Gates G1-G4 and G11 (grey) are not used in this work.

**Figure 2.** (a) Measured charge stability diagram of the charge sensor QD in a similar device as shown in Figure 1(a). The QD is defined by setting $V_{g1}$=−0.95 V and $V_{g3}$=−0.35 V and is tuned by gate voltage $V_{g2}$. (b) Measured charge stability diagram of a single QD defined in the MQD nanowire of the device using gates G7 and G9 as barrier gates (with $V_{G7}$=−0.44 V and $V_{G9}$=−0.32 V) and G8 as a plunger gate. (c) Source-drain current $I_{sd}$ of the charge sensor as a function of $V_{g2}$ measured at source-drain voltage $V_{sd}$=0.1 mV. The red star marks the position at which some charge detection measurements are performed in this work. (d) Simultaneous measurements of the current through the charge sensor $I_{sd}$ at $V_{sd}$=0.1 mV and $V_{g2}$=0.114 V [i.e., at the declining slope of a Coulomb current peak as indicated by the red star in (c)] and through the single QD $I_{SD}$ in the MQD nanowire at $V_{SD}$=0.2 mV as a function of $V_{G8}$. Distinct jumps in $I_{sd}$ appear when $I_{SD}$ is at Coulomb current peaks.

**Figure 3.** (a) Source-drain current $I_{SD}$ of a TQD measured as a function of $V_{G7}$ and $V_{G9}$ (charge stability diagram) at $V_{SD}$=70 μV. The TQD studied in this work is defined and manipulated using gate G5-G10 (with setting $V_{G5}$=−0.31 V, $V_{G6}$=−0.072 V, $V_{G8}$=−0.42 V, and $V_{G10}$=−0.54 V). Three groups of current lines of different slopes arise from resonant transport through QD1 (vertical), QD2 (inclined), and QD3 (horizontal) in the TQD. (b) Schematic view of energy level alignments for the cases when only QD2 (upper panel) or only QD1 (lower panel) is on resonance with the Fermi level of the source and drain reservoirs. (c) and (d) Charge stability diagrams of the TQD measured as the transconductance $dI_{sd}/dV_{G9}$ of the charge sensor at $V_{sd}$=0.2 mV as a function of $V_{G7}$ and $V_{G9}$, by setting the sensor at the rising slope and the declining slope [as indicated by the red star in Fig. 2(c)] of a Coulomb current peak.

**Figure 4.** (a)-(c) Charge stability diagrams of the TQD measured as the transconductance $dI_{sd}/dV_{G9}$ of the charge sensor at $V_{sd}$=0.2 mV and at the declining slope of a Coulomb current peak, as indicated by the red star in Fig. 2(c), as a function of $V_{G7}$ and $V_{G9}$ at different $V_{G6}$. The dashed circles in (a) and (c) mark the regions with each containing two triple points. The dashed circle in (b) marks the region where a QP and several triple points could be present. (d)-(f) Schematics for the energy level alignments of the TQD in the regions marked by the dashed circles in (a)-(c), respectively, without taking the couplings between individual QD levels into account.

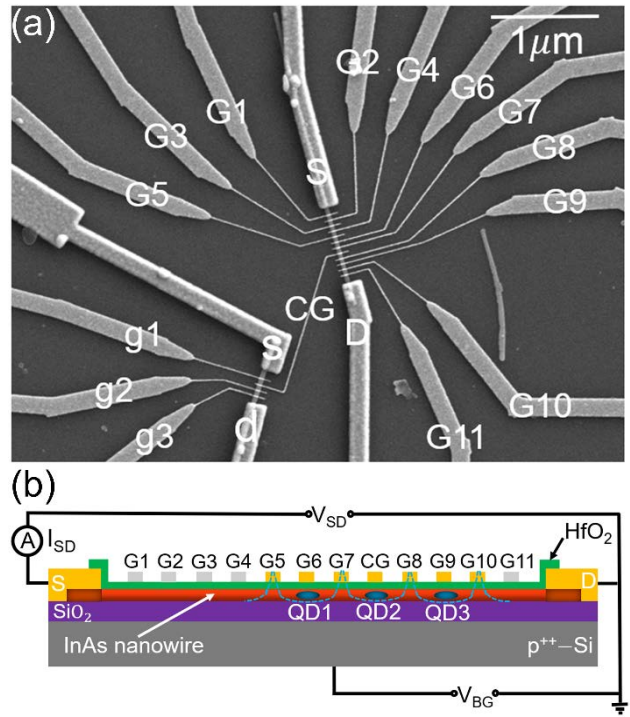

Figure 1, Weijie Li *et al*.

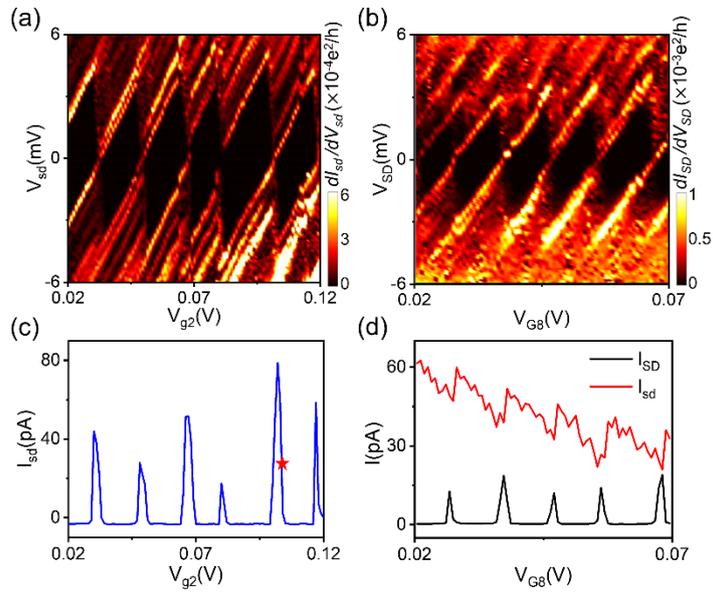

Figure 2, Weijie Li *et al*.

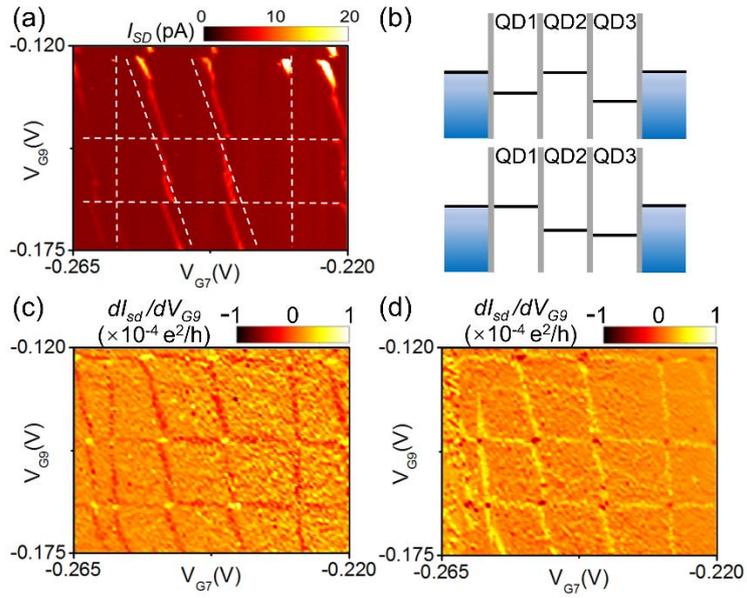

Figure 3, Weijie Li *et al*.

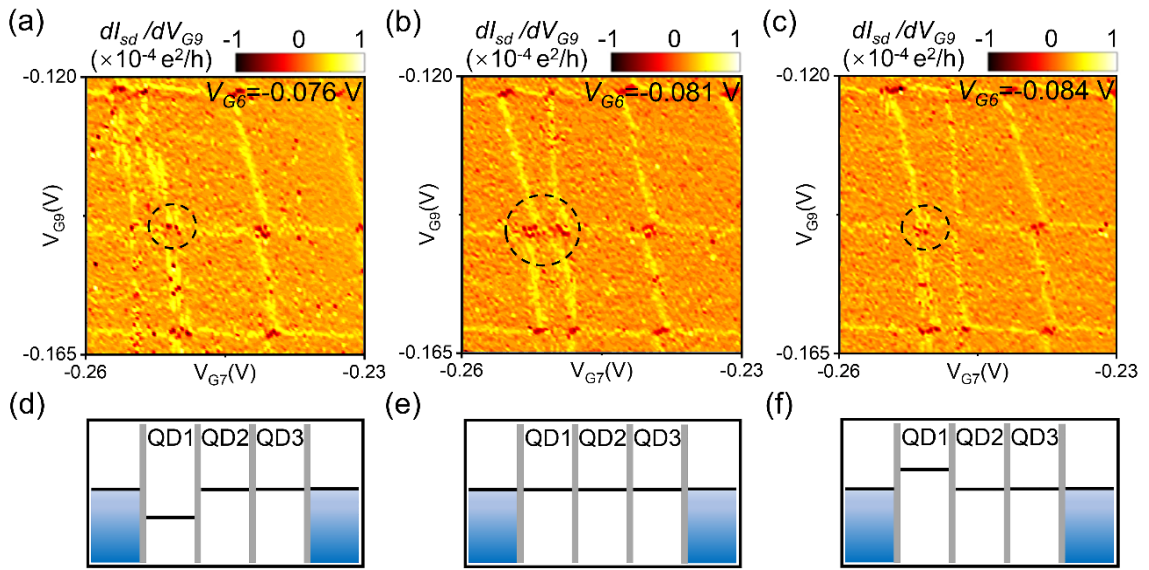

Figure 4, Weijie Li *et al*.

**Supplementary Material for**

**Detection of charge states of an InAs nanowire triple quantum dot with an integrated nanowire charge sensor**

Weijie Li,[1,2] Jingwei Mu,[1,2] Shaoyun Huang,[1] Dong Pan,[3] Jianhua Zhao,[3,4] and H. Q. Xu[1,4,a)]

[1] Beijing Key Laboratory of Quantum Devices, Key Laboratory for the Physics and Chemistry of Nanodevices and Department of Electronics, Peking University, Beijing 100871, China

[2] Academy for Advanced Interdisciplinary Studies, Peking University, Beijing 100871, China

[3] State Key Laboratory of Superlattices and Microstructures, Institute of Semiconductors, Chinese Academy of Sciences, P.O. Box912, Beijing 100083, China

[4] Beijing Academy of Quantum Information Sciences, Beijing 100193, China

a) Electronic address: hqxu@pku.edu.cn

(Dated: December 10, 2020)

**Tuning to a region where all the three QDs in the TQD are close to be on resonance with the electrodes in a slightly stronger coupling condition**

We have tuned our TQD device to a region where all the three QDs in the device are close to be on resonance with the electrodes in a slightly stronger coupling condition. The experimental results obtained with a similar tuning process as in Figure 4 of the main article are shown in Figure S1. Here, the voltages applied to barrier gates G5, G8 and G10 were set at $V_{G5}=-0.305$ V, $V_{G8}=-0.4$ V, $V_{G10}=-0.53$ V, which are less negative compared to the corresponding voltage values of $V_{G5}=-0.31$ V, $V_{G8}=-0.42$ V, $V_{G10}=-0.54$ V used in Figure 4 of the main article. The gate voltage $V_{G7}$ is also swept in a less negative range of $-0.173$ V to $-0.133$ V, compared to that of $-0.26$ V to $-0.23$ V in Figure 4 of the main article. Figures S1(a)-S1(c) display the charge stability diagrams obtained via direct transport measurements of the TQD at three different $V_{G6}$. Finite current lines of three different slopes (i.e., nearly vertical, inclined, and nearly horizontal lines) can now be observed more clearly in Figures S1(a)-S1(c). The current lines with an inclined slope [see, e.g., the current lines marked by inclined pink dashed lines in Figure S1(b)] can be assigned to resonant transport through energy levels of QD2. The nearly vertical current lines [see, e.g., the current lines marked by nearly vertical green dashed lines in Figure S1(b)] arise from resonant transport through energy levels of QD1. Similarly, the nearly horizontal current lines [see e.g., the current lines marked by nearly horizontal white dashed lines in Figure S1(b)] arise from resonant transport through energy levels of QD3. Here, as the voltages applied to all the barrier gates are less negative compared to that in Figure 4 of the main article, the couplings between the three QDs are stronger and the differences between the slopes of the resonant current lines of the three QDs are smaller. Figures S1(d)-S1(f) show the charge stability diagram of the TQD measured at the three different $V_{G6}$ via the nanowire QD charge sensor which is defined in the same way as in the main article with $V_{g2}$ being again set at $V_{g2}=0.114$ V [i.e., at the declining slope of the same Coulomb current peak as marked by the red star in Figure 2(c) of the main article]. Here, in Figures S1(d)-S1(f), it is more clearly seen that as $V_{G6}$ is tuned to bring an energy level of QD1 very close to be on resonance with the energy levels of QD2 and QD3, charge state transition lines of the three QDs in the close-to-be on-resonance region show complex evolutions. The black dashed circle in Figure S1(e) marks the region where the energy levels of all the three QDs in the TQD are very close to be on resonance. Kinks in the charge state transition lines in the region marked by the black dashed circle are observed. These kinks are signs of the formation of a QP and several triple points in the TQD device.

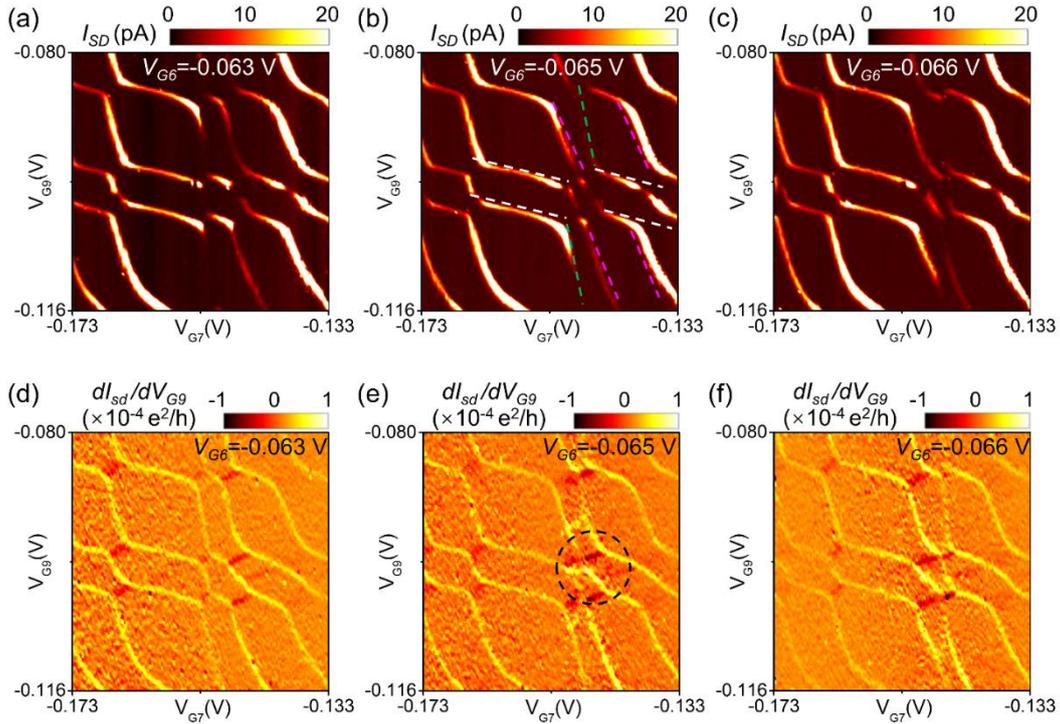

**Figure S1.** (a)-(c) Charge stability diagrams measured as the source-drain current $I_{SD}$ of the TQD at $V_{SD}$=70 μV as a function of $V_{G7}$ and $V_{G9}$ at three different $V_{G6}$. The TQD studied here is defined and manipulated using gates G5-G10 with $V_{G5}$, $V_{G8}$, and $V_{G10}$ being fixed at $V_{G5}$=−0.305 V, $V_{G8}$=−0.4 V and $V_{G10}$=−0.53 V, but with $V_{G6}$, $V_{G7}$, and $V_{G9}$ being varied as shown in the three panels. (d)-(f) Corresponding charge stability diagrams of the TQD measured as the transconductance $dI_{sd}/dV_{G9}$ of the QD charge sensor at $V_{sd}$=0.2 mV as a function of $V_{G7}$ and $V_{G9}$. Here, the QD charge sensor is defined exactly as in the main article with setting $V_{g2}$=0.114 V [i.e., at the declining slope of the same Coulomb current peak as marked by the red star in Figure 2(c) of the main article] The gate voltages that are used to define and manipulate the TQD are the same as in panels (a)-(c). The black dashed circle in (e) marks a region in which all the three QDs in the TQD are very close to be on resonance with the source and drain electrodes, and kinks in the charge state transition lines (i.e., the signs of the formation of a QP and several triple points) are visible.